\documentclass[%
reprint,%
amssymb, amsmath,%
aip,cha,%
]{revtex4-1}

\usepackage{bm}%
\usepackage[colorlinks=true,linkcolor=blue, citecolor=blue]{hyperref}
\usepackage{graphicx}
\usepackage{multirow}
\usepackage{subfigure}
\usepackage{color,soul}
\usepackage{footnote}
\expandafter\ifx\csname package@font\endcsname\relax\else
\expandafter\expandafter
\expandafter\usepackage
\expandafter\expandafter
\expandafter{\csname package@font\endcsname}%
\fi
\begin{document}
    
    \title{p-Bits for Probabilistic Spin Logic}%
    
    \author{Kerem Y. Camsari}
    \affiliation{School of Electrical and Computer Engineering, Purdue University, West Lafayette, IN 47907, USA}
    \author{Brian M. Sutton}
    \affiliation{School of Electrical and Computer Engineering, Purdue University, West Lafayette, IN 47907, USA}
    \author{Supriyo Datta}
     \affiliation{School of Electrical and Computer Engineering, Purdue University, West Lafayette, IN 47907, USA}
    
    \date{\today}%
    
    \begin{abstract}
    We introduce the concept of a probabilistic or p-bit, intermediate between the standard bits of digital electronics and the emerging q-bits of quantum computing. We show that low barrier magnets or LBM's provide a natural physical representation for p-bits and can be built either from perpendicular magnets (PMA) designed to be close to the in-plane transition or from circular in-plane magnets (IMA). Magnetic tunnel junctions (MTJ) built using LBM's as free layers can be combined with standard NMOS transistors to provide three-terminal building blocks for large scale probabilistic circuits that can be designed to perform useful functions. Interestingly, this three-terminal unit looks just like the 1T/MTJ device used in embedded MRAM technology, with only one difference: the use of an LBM for the MTJ free layer. We hope that the concept of p-bits and p-circuits will help open up new application spaces for this emerging technology. However, a p-bit need not involve an MTJ, any fluctuating resistor could be combined with a transistor to implement it, while completely digital implementations using conventional CMOS technology are also possible. The p-bit also provides a conceptual bridge between two active but disjoint fields of research, namely stochastic machine learning and quantum computing. First, there are the applications that are based on the similarity of a p-bit to the binary stochastic neuron (BSN), a well-known concept in machine learning. Three-terminal p-bits could provide an efficient hardware accelerator for the BSN. Second, there are the applications that are based on the p-bit being like a poor man's q-bit. Initial demonstrations based on full SPICE simulations show that several optimization problems including quantum annealing are amenable to p-bit implementations which can be scaled up at room temperature using existing technology.

    \end{abstract}
    
    \maketitle
    \tableofcontents
   
 
  \section{Introduction} 
  
   \subsection{Between a bit and a q-bit} 

\begin{figure} [t!]
        \includegraphics[width=0.85\linewidth]{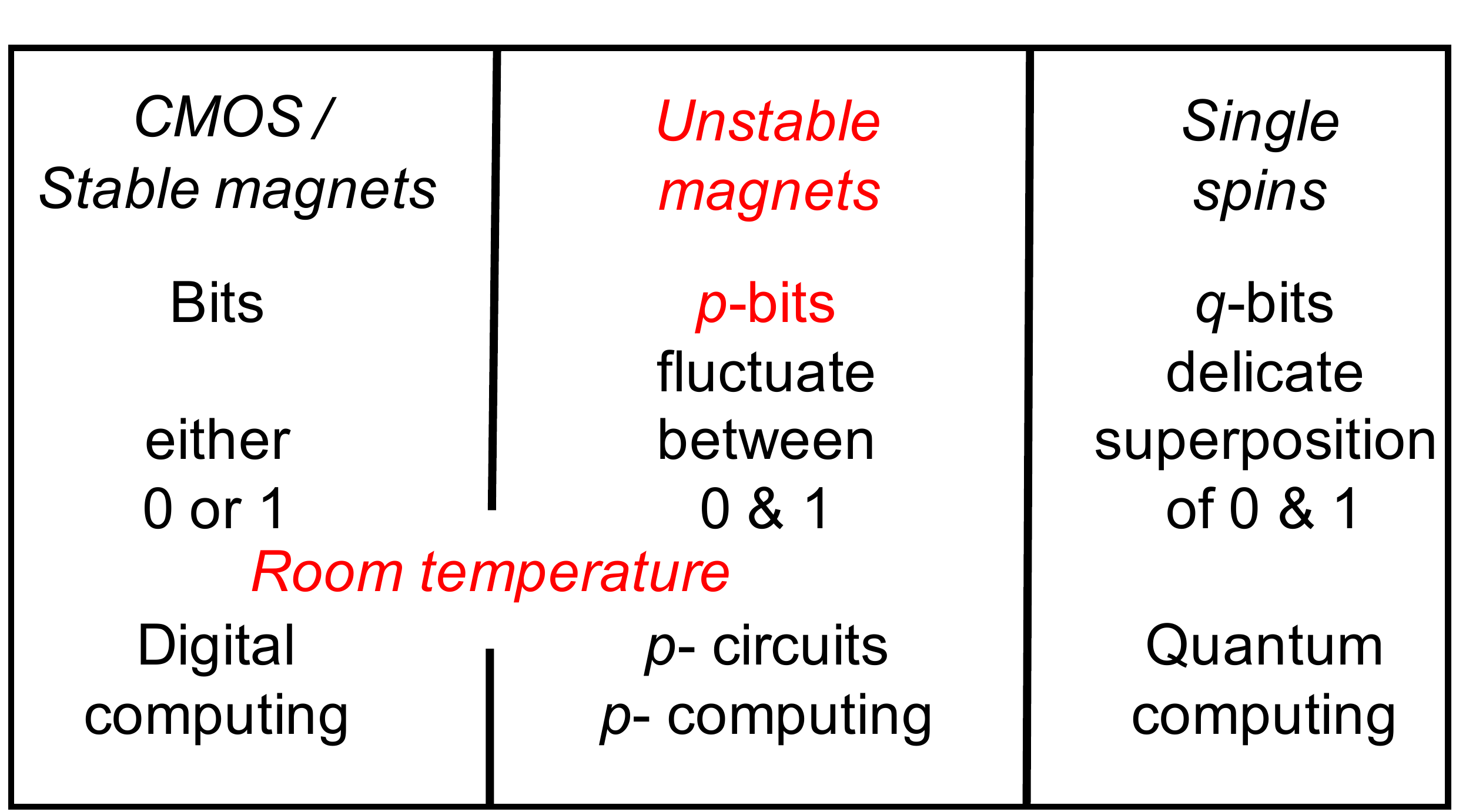}
        \caption{\label{bpq} {\bf Between a bit and a q-bit: The p-bit} Digital computers use deterministic strings of 0's and 1's called bits to represent information in a binary code. The emerging field of quantum computing is based on q-bits representing a delicate superposition of 0 and 1 that typically requires cryogenic temperatures. We envision a class of probabilistic computers or p-computers operating robustly at room temperature with existing technology based on p-bits which are classical entities fluctuating rapidly between 0 and 1. Although spins provide a nice unifying paradigm for illustrating the transition from bits to p-bits and q-bits, it should be noted that the physical realization of a p-bit need not involve spins or spintronics; non-spintronic implementations can be just as feasible. }
     \label{fi:fig1}
    \end{figure}

    Modern digital circuits are based on binary bits that can take on one of two values, 0 and 1, and are  stored using well-developed technologies at room temperature. At the other extreme are quantum circuits based on q-bits which are delicate superpositions of 0 and 1 requiring the development of novel technologies typically working at cryogenic temperatures. This article is about what we call probabilistic bits or $\textit{p-bits}$ that are classical entities fluctuating rapidly between 0 and 1. We will argue that we can use existing technology to build what we call \textit{p-circuits} that should function robustly at room temperature while addressing some of the applications commonly associated with quantum circuits (Fig.~\ref{fi:fig1}).
        
     How would we represent a p-bit physically? Let us first consider the two extremes, namely the bit and the q-bit. A q-bit is often represented by the spin of an electron, while a bit is often represented by binary voltage levels in digital elements like flip-flops and floating-gate transistors. However, bits can also be represented by magnets \cite{chen_advances_2010} which are basically collections of a very large number of spins. In a magnet, internal interactions make the energy a minimum when the spins all point either parallel or anti-parallel to a specific direction, called the easy axis. These two directions represent  0 and 1 and are separated by an energy barrier, $E_b$, that ensures their stability.
     
     How large is the barrier? A nanomagnet flips back and forth between 0 and 1 at a rate determined by the energy barrier:   $   \tau \sim \tau_0 \ \exp (E_b/k_BT)$  where $\tau_0$  typically has a value between picoseconds and nanoseconds\cite{lopez2002transition}. Assuming a $\tau_0 $ of a nanosecond, a barrier of $E_b \sim 40 \  k_BT$, for example, would retain a 0 (or a 1) for $\sim$ $\text{10 years}$, making it suitable for long term memory while a smaller barrier of $E_b \sim 14 \  k_BT$, would only ensure a short term memory $\sim$ $\text{1 ms}$\cite{locatelli2014noise}.
     
     It has been recognized that this stability problem also represents an opportunity. Unstable low barrier magnets (LBM) could be used to implement useful functions like random number generation (RNG) \cite{majetich_2018,vodenicarevic_low-energy_2017,vodenicarevic_circuit-level_2018} by sensing the randomly fluctuating magnetization to provide a random time varying voltage. With such applications in mind, we would want magnets to have as low a barrier as possible, so that many random numbers are generated in a given amount of time. Indeed a ``zero" barrier magnet with $E_b \leq k_B T$ flipping back and forth in less than a nanosecond would be ideal.
     
     How can we reduce the energy barrier? Since $E_b = H_K M_s \Omega / 2$, the basic approach is to reduce the total magnetic moment by reducing volume $\Omega$, and/or engineer a small anisotropy field $H_K$ \cite{debashis2018designing}. This can be done with perpendicular magnets (PMA) designed to be close to the in-plane transition. A less challenging approach seems to be to use circular in-plane magnets (IMA) \cite{cowburn_single-domain_1999,debashis_experimental_2016,debashis2018designing}. We will refer to all these possibilities collectively as LBM's as opposed to say superparamagnets which have more specific connotations in different contexts\cite{sutton_intrinsic_2017,faria2017low,camsari_stochastic_2017, camsari_implementing_2017, mizrahi2018neural,bapna2017current,locatelli2014noise}.

     We could use LBM's to represent the  probabilistic bits or $\textit{p-bits}$ that we alluded to. We have argued that if these \emph{p-bits}  can be incorporated into proper transistor-like structures with gain, then the resulting $\textit{three-terminal p-bits}$ could be interconnected to build \emph{p-circuits} that perform useful functions \cite{behin-aein_building_2016, sutton_intrinsic_2017, camsari_stochastic_2017}, not unlike the way transistors are interconnected to build useful digital circuits. However, unlike digital circuits these probabilistic p-circuits incorporate features reminiscent of quantum circuits.

This connection was nicely articulated by Feynman in a seminal paper \cite{feynman_simulating_1982}, where he described a quantum computer that could provide an efficient simulation of quantum many-body problems. But to set the stage for quantum computers, he first described a probabilistic \emph{p-computer} which could efficiently simulate classical many-body problems:

\begin{quote}  
\ldots \textit{``the other way to simulate a probabilistic nature, which I'll call N \ldots is by a computer C which itself is probabilistic, \ldots in which the output is not a unique function of the input \ldots it simulates nature in this sense: that C goes from some \ldots initial state \ldots to some final state with the same probability that N goes from the corresponding initial state to the corresponding final state \ldots If you repeat the same experiment in the computer a large number of times \ldots it will give the frequency of a given final state proportional to the number of times, with approximately the same rate \ldots as it happens in nature.''}
\end{quote}
\noindent There are many practical problems of great interest which involve large networks of probabilistic quantities. These problems should be simulated efficiently by p-computers of the type envisioned by Feynman. Our purpose here is to discuss appropriate $\textit{hardware building blocks}$ that can be used to $\textit{build}$ them  \cite{behin-aein_building_2016} and possible applications they could be used for. In this context, let us note that although spins provide a nice unifying paradigm for illustrating the transition from bits to p-bits and q-bits, the physical realization of a p-bit need not involve spins or spintronics; non-spintronic implementations can be just as feasible.

 \subsection{Binary stochastic neuron (BSN)}

Interestingly the concept of a p-bit connects naturally to another concept well-known in the field of machine learning, namely that of a binary stochastic neuron (BSN) \cite{ackley_learning_1985,neal1992connectionist} whose response $m_i$   to an input $I_i$ can be described mathematically by 
 \begin{align}
   m_i = {\rm{sgn}}[ \mathrm{tanh} \ {I_i} - r ] 
   \label{eq:pbit}
\end{align}
\noindent where r is a  random number uniformly distributed between $-$1 and +1\footnote{Eq.~\ref{eq:pbit} can be equivalently written as  $m_i = {\rm{sgn}}[ \mathrm{tanh} \ {I_i} +r ]$.}. Here we are using bipolar variables $m_i=\pm1$ to represent the \emph{0} and \emph{1} states. If we use binary variables $m_i=0,1$ the corresponding equation would look different\footnote{The signum function (sgn) would be replaced by the step function ($\Theta$) and the tanh function would be replaced by the sigmoid function ($\sigma$) such that   $m_i = \Theta[\sigma(2 I_i) - r_0]$ where the random number $r_0$ is uniformly distributed between 0 and 1.}. When combined with a synaptic function described by
 \begin{align}
   {I_i} = \sum_j{W_{ij} \ m_j}
   \label{eq:synapse}
\end{align}
\noindent we have a probabilistic network that can be designed to perform a wide variety of functions through a proper choice of the weights, $W_{ij}$. A separate bias term $h_i$ is often included in Eq.~\ref{eq:synapse} but we will not write it explicitly, assuming that it is included as the weighted input from an extra p-bit that is always +1.

Eqs.~\ref{eq:pbit} and \ref{eq:synapse} are widely used in many modern algorithms but they are commonly implemented in software. Much work has gone into developing suitable hardware accelerators for matrix multiplication of the type described by Eq.~\ref{eq:synapse} (See for example, Ref.~\cite{hu2016dot}). Three-terminal p-bits would provide a $\textit{hardware accelerator}$ for Eq.~\ref{eq:pbit}. Together they would function like a probabilistic computer.

\begin{figure} [t!]
        \includegraphics[width=0.85\linewidth]{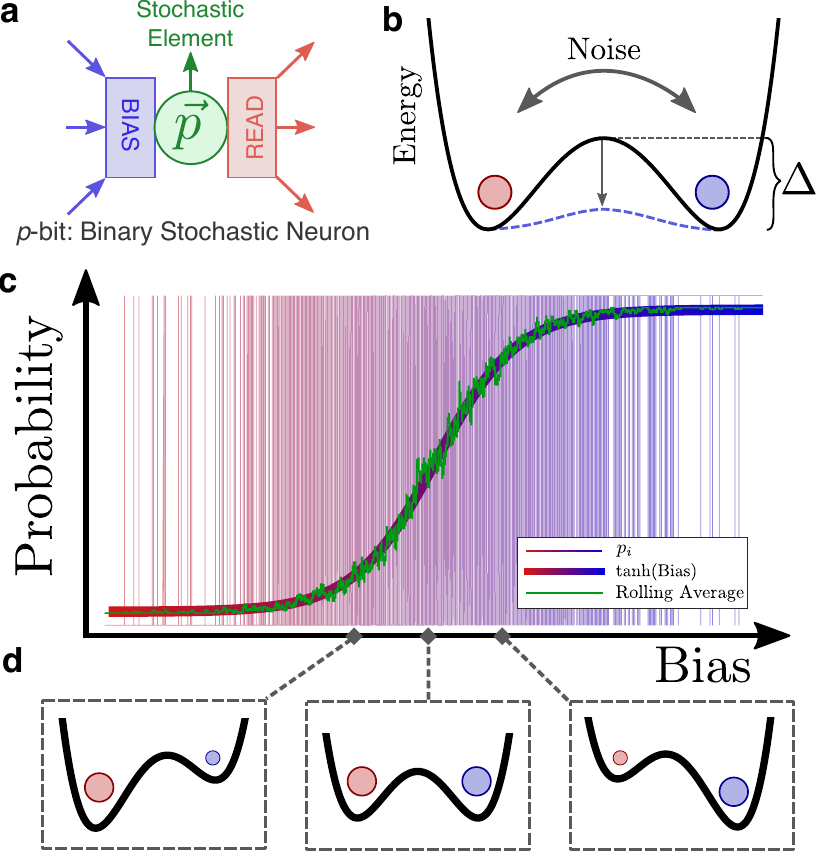}
        \caption{\label{fig:pbit_tunable_rng} {\bf Three terminal p-bit:} {\bf a.}  A hardware implementation of the BSN (Eq.~(\ref{eq:pbit})) requires a central stochastic element with input and output terminals that provide the ability to read and bias the element. {\bf b.} The stochastic element can be visualized as  going back and forth between two low energy states at a rate that depends exponentially on the barrier $E_b$ that separates them: $\tau = \tau_0 \exp(\Delta/k_\text{B}T)$ {\bf c.} The bias terminal adjusts the relative energies of the two states thereby controlling the probabilities of finding the element in the two states.}
             \label{fi:fig2}
    \end{figure}

Note that a hardware accelerator for Eq.~(\ref{eq:pbit})  requires more than just an RNG. We need a \emph{tunable} RNG whose output $m_i$ can be biased through the input terminal $I_i$ as shown in Fig.~\ref{fi:fig2}. Two distinct designs for a three-terminal p-bit have been described \cite{camsari_stochastic_2017,camsari_implementing_2017} both of which use a magnetic tunnel junction (MTJ), a popular ``spintronic'' device used in magnetic random access memory (MRAM)\cite{bhatti2017spintronics}. However, MRAM applications use stable MTJ's that can store information for many years, while a p-bit makes use of \emph{``bad"} MTJ's with low barriers. 

The LBM-based implementation of the BSN described here is conceptually very different from a \emph{clocked} approach using stable magnets where a stochastic output is obtained every time a clock pulse is applied \cite{behin2012modeling,behin2014computing,choi2014magnetic,fukushima2014spin,querlioz_2015,behin-aein_building_2016,sengupta2016probabilistic,lv2017single}. All of these approaches work with stable magnets, although LBM's could be used to reduce the switching power that is needed.

In this paper we will focus on $\textit{unclocked, asynchronous}$ operation using LBM-based hardware accelerators for the BSN (Eq.~(\ref{eq:pbit})) \cite{faria2017low,sutton_intrinsic_2017,camsari_stochastic_2017}.  But can an asynchronous circuit provide the \textit{sequential} updating of the BSN's described by Eq.~(\ref{eq:pbit}) that is required for Gibbs sampling\cite{geman1984stochastic} and is commonly enforced in software through a \emph{for loop}? The answer is ``yes" as shown both in SPICE simulations \cite{sutton_intrinsic_2017} as well as arduino-based emulations \cite{pervaiz2018weighted}, \emph{provided} the synaptic function in Eq.~(\ref{eq:synapse}) has a delay that is less than or comparable to the response time of the BSN, Eq.~(\ref{eq:pbit}).

It should be noted that unclocked operation is a rarity in the digital world and most applications will probably use a clocked, sequential approach with dedicated sequencers that update connected p-bits sequentially. A fully digital implementation of p-circuits using such dedicated sequencers has been realized in Ref.~\cite{pervaiz2018weighted}.  Synchronous operation can be particularly useful if synaptic delays are large enough to interfere with natural asynchronous operation.

Here, we focus on unclocked operation in order to bring out the role of a p-bit in providing a conceptual bridge between two very active fields of research, namely \emph{stochastic machine learning} and \emph{quantum computing}. On the one hand p-bits could provide a hardware accelerator for the BSN (Eq.~(\ref{eq:pbit})) thereby enabling applications inspired by machine learning (\textbf{Section III}). On the other hand, p-bits are the classical analogs of q-bits: robust room temperature entities accessible with current technology that could enable at least some of the applications inspired by quantum computing (\textbf{Section IV}). But before we discuss applications, let us briefly discuss possible hardware approaches to implementing p-bits (\textbf{Section II}).


\section{Hardware Implementation}

 \subsection{Three-terminal p-Bit} 
 
RNG's represent an important component of modern electronics and have been implemented using many different approaches, including Johnson-Nyquist noise of resistors \cite{cheemalavagu_probabilistic_2005}, phase noise of ring oscillators \cite{bucci_high-speed_2003}, process variations of SRAM cells \cite{holcomb_power-up_2009} and other physical mechanisms. However, as noted earlier, we need what appears to be a completely new 3-terminal device whose input $I_i$ biases its stochastic output $m_i$ as shown in Fig.~\ref{fi:fig2}c.

\begin{figure} [!t]
    \includegraphics[width=0.95\linewidth]{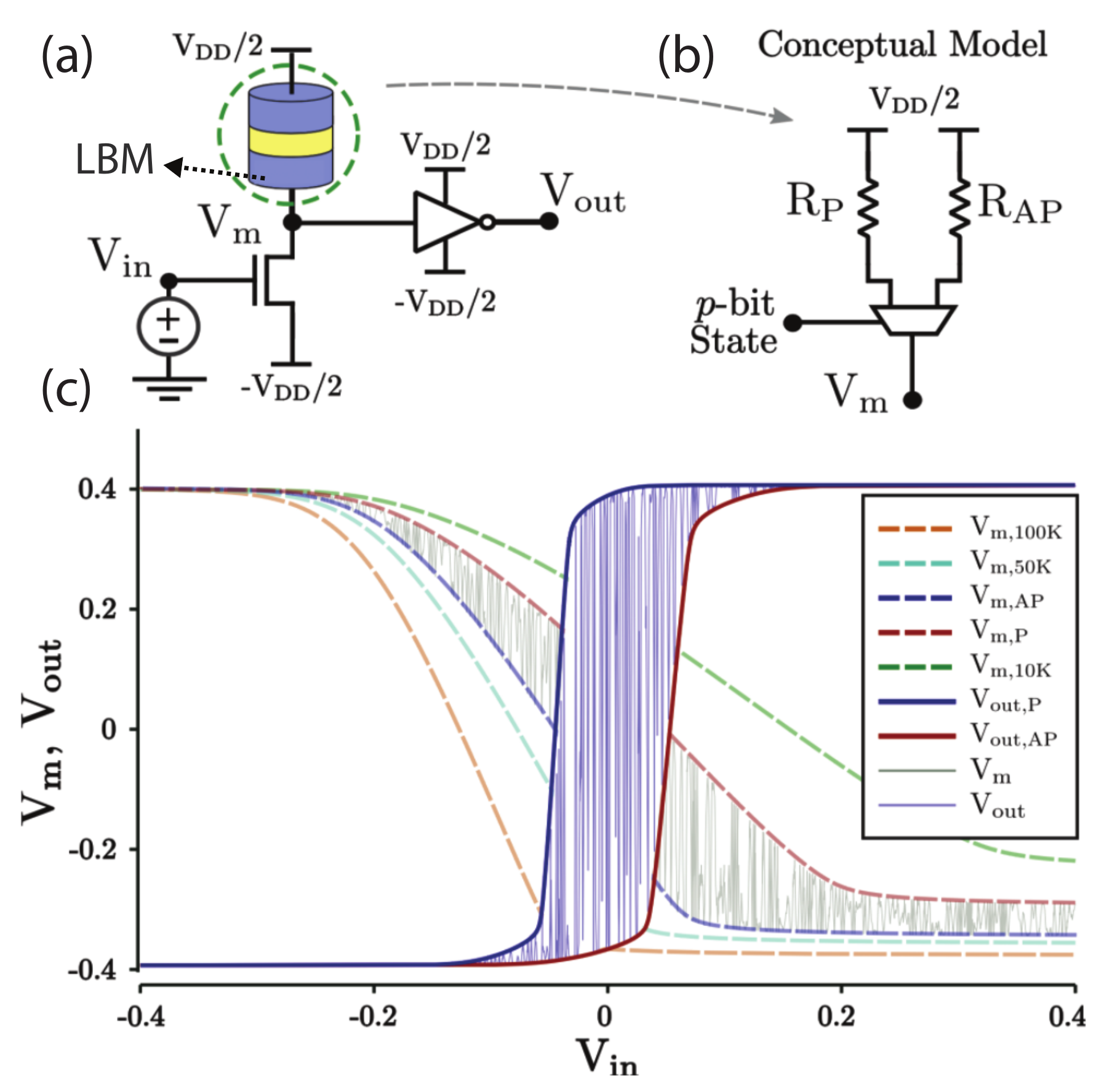}
    \caption{\label{fig:embedded_mram} {\bf Embedded MRAM p-bit:} {\bf a.} An NMOS pull-down transistor in series with a stochastic-MTJ whose resistance fluctuates between $RP$ and $R_{AP}$ as shown in {\bf b.} {\bf c.} Using a 14 nm HP-FinFET model \cite{cao2002predictive} the input voltage, $V_\text{in}$, versus mid-point, $V_m$, and output $V_\text{out}$, voltages is simulated in SPICE. Several fixed resistances are shown to convey how $V_m$ would vary with modifications to the parallel and anti-parallel resistances.}
    \label{fi:fig3}
 \end{figure}

A recent paper \cite{camsari_implementing_2017} shows that such a 3-terminal tunable RNG can be built simply by combining a 2-terminal fluctuating resistance with a transistor (Fig.~\ref{fi:fig3}). This seems very attractive at least in the short run, since the basic structure (Fig.~\ref{fi:fig3}a) closely resembles the 1T/MTJ structure commonly used for MRAM applications. The first modification that is required is to replace the stable free layer of the MTJ with an LBM. The second modification is to add an inverter to the drain output that amplifies the fluctuations caused by the MTJ resistance. 

An MTJ is a device with two magnetic contacts whose electrical resistance $R_{MTJ}$ takes on one of two values $R_P$ and $R_{AP}$ depending on whether the magnets are parallel (P) or antiparallel (AP). MTJs are typically used as memory devices, though in recent years applications of MTJs for logic and novel types of computation have been discussed \cite{wang2005programmable,matsunaga2008fabrication,ohno2010magnetic,lyle2011magnetic,yao2012magnetic,grollier2016spintronic,locatelli2014spin}.  

Standard MTJ devices go to great lengths to ensure that the magnets they use are stable and can store information for many years. The resistance of $\textit{bad}$ MTJ's, on the other hand, constantly fluctuates between $R_{P}$ and $R_{AP}$ \cite{locatelli2014noise}. If we put it in series with a transistor which is a voltage controlled resistance $R_T(V_{in})$ then the voltage $V_m$ (Fig.~\ref{fi:fig3}) can be written as
 \begin{align*}
   {V_m} = \frac{V_{DD}} {2} \frac{R_T(V_{in})- R_{MTJ}}{R_T (V_{in})+ R_{MTJ}}
\end{align*}

The magnitude of this fluctuating voltage $V_m$ is largest when the transistor resistance $R_T \sim R_P$ or $R_{AP}$ but gets suppressed if $R_T \ll R_P$ or if $R_T \gg R_{AP}$. The input voltage controls $R_T$ thereby tuning the stochastic output $V_m$ as shown in Fig.~\ref{fi:fig3}c. It was shown that an additional inverter provides an output that is approximately described by an expression that looks just like the BSN (Eq.~\ref{eq:pbit})
 \begin{equation}
 \overbrace{\frac{V_{out,i}}{V_{DD}/2}}^{m_i} \approx \hspace{0.1in} {\rm{sgn}}\bigg{\{} \mathrm{tanh} \overbrace{\frac{V_{in,i}}{V_0}}^{I_i} - r \bigg{\}} 
 \label{eq:scaling}
\end{equation}
but with dimensionless variables like $m_i$ and $I_i$ replaced by scaled circuit voltages $V_{out}$ and $V_{in}$.

The scheme in Fig.~\ref{fi:fig3} provides tunability through the series transistor and does not involve the physics of the fluctuating resistor. Ideally, the magnet is unaffected by the change in the transistor resistance though the drain current, in principle, could pin the magnet. In our simulations that are based on Ref.~\cite{camsari_implementing_2017}, we take the pinning current into account through a spin-polarized current ($I_s$) proportional to an effective  fixed layer polarization and the drain current $(I_D)$, $\vec{I}_s = (P) I_{D} \hat x$, where $\hat x$ is the fixed layer direction. This spin-current enters the sLLG equation that calculates an instantaneous magnetization which in turn controls the MTJ resistance. 

We note that any significant pinning around zero input voltage $V_{in,i}$ has to be minimized through proper design, especially for low barrier perpendicular magnets which are relatively easy to pin. Unintentional pinning \cite{liyanagedera2017stochastic} should in general not be an issue for circular in-plane LBM's due to the strong demagnetizing field. The pinning behavior for the average (steady-state) magnetization can be qualitatively understood by numerical simulations of the sLLG equation. In the case of low-barrier perpendicular magnets the spin-torque pinning needs to overcome the thermal noise and therefore the pinning current is of order $I_{PMA} \approx 2 (q /\hbar) \alpha kT $ where $\alpha$ is the damping coefficient of the magnet. In the case of circular in-plane magnets, the pinning current  is of order $I_{IMA} \approx 2 (q /\hbar) \alpha H_D M_s \mathrm{Vol.}$, which is much larger than $I_{PMA}$ since for for typical parameters ($H_D M_s \mathrm{Vol.}\gg kT$).

Since the state of the magnet is not affected, if the input voltage $V_{in,i}$ in Eq.~\ref{eq:scaling} is changed at t=0, the statistics of the output voltage $V_{out,i}$  will respond within tens of picoseconds (typical transistor switching speeds) \cite{nikonov2015benchmarking} irrespective of the fluctuation rates of the magnet. However, the magnet fluctuations will determine the correlation time of the random number \emph{r} in Eq.~\ref{eq:scaling}.

Alternatively one can envision structures where the input controls the statistics of the fluctuating resistor itself, through phenomena such as the spin-Hall effect \cite{camsari_stochastic_2017} or the magneto-electric effect \cite{camsari2018equivalent} based on a voltage control of magnetism (see for example \cite{biswas2017experimental, manipatruni2018beyond}). In that case, both the speed of response and the correlation time of the random number \emph{r} will be determined by the specific phenomenon involved.

\textit{Non-spintronic implementations:} Note that the structure in Fig.~\ref{fi:fig3} could use any fluctuating resistor including CMOS-based units in place of the MTJ showing that the physical realization of a p-bit need not involve spins\cite{jerry2017random}. For example, a linear feedback shift register (LFSR) is often used to generate a pseudo-randomly fluctuating bit stream\cite{lewis_generalized_1973}. We can apply this fluctuating voltage to the gate of a transistor to obtain a fluctuating resistor which can replace the MTJ in Fig.~\ref{fi:fig3}a. We note that the main appeal of the structure in Fig.~\ref{fi:fig3} lies in its simplicity, since a 1T/1MTJ design coupled with two more transistors provide the tunable randomness in a compact transistor-like building block. Using completely digital p-circuit implementations \cite{pervaiz2018weighted} could offer short term scalability and reliability but they would consume a much larger area and power per p-bit. 

\subsection{Weighted p-bit}

\begin{figure} [t!]
        \includegraphics[width=0.85\linewidth]{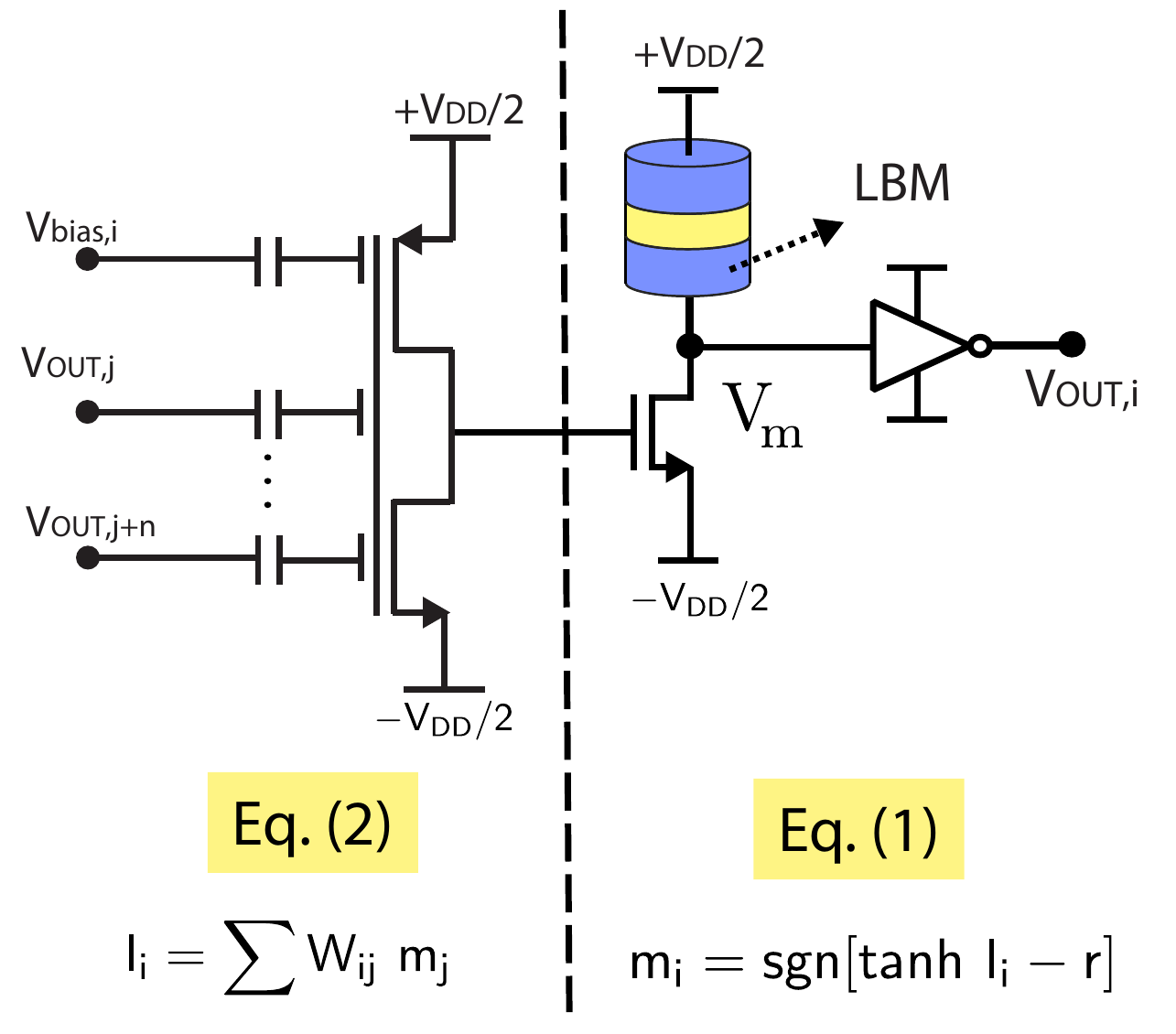}
        \caption{\label{fig:fig8_orchi_weight} {\bf Example of a weighted p-bit integrating relevant parts of the synapse onto the neurons: } Leveraging floating-gate devices along the lines proposed in neuMOS \cite{shibata_functional_1992} devices, a collection of synapse inputs (from 1 to n)  can be summed to produce the bias voltage, $V_{\text{IN},i}$  for a voltage driven p-bit\cite{hassan_voltage-driven_2018}. }
        \label{fi:fig4}
\end{figure}

The structure in Fig.~\ref{fi:fig3} gives us a ``neuron'' that implements Eq.~\ref{eq:pbit} in hardware. Such neurons have to be used in conjunction with a ``synapse'' that implements Eq.~\ref{eq:synapse}. 
Alternatively we could design a ``weighted p-bit'' that integrates each element of Eq.~\ref{eq:pbit} with the relevant part of Eq.~\ref{eq:synapse}. For example, we could use floating gate devices along the lines proposed in neuMOS \cite{shibata_functional_1992} devices as shown in Fig.~\ref{fi:fig4}. From charge conservation we can write
\begin{align*}
          \sum_j (V_{out,j} - V_{in,i}) C_{i,j} - V_{in,i} C_0 = 0
\end{align*}
\noindent where $C_0$ is the input capacitance of the transistor. This can be rewritten as
\begin{align*}
V_{in,i} =  \sum_j  \frac{C_{i,j}}{C_0+ \sum_j C_{i,j}} \ \ V_{out,j} 
\label{eq:hw_synapse}
\end{align*}
\noindent By scaling $V_{in}$ and $V_{out}$ (see Eq.~\ref{eq:scaling}) to play the roles of the dimensionless quantities $I_i$ and $m_i$ respectively, we can recast Eq.~\ref{eq:hw_synapse} in a form similar to Eq.~\ref{eq:synapse}:
\begin{align}
\overbrace{\frac{V_{in,i}}{V_0} }^{I_i} =  \sum_j  \overbrace{\frac{V_{DD}}{2 V_{0}} \frac{C_{i,j}}{C_0+ \sum_j C_{i,j}}}^{W_{ij}} \ \   \overbrace{\frac{V_{out,j}}{V_{DD}/2}}^{m_i}
\end{align}

The weights $W_{ij}$ can be adjusted by controlling the specific capacitors $C_{ij}$ that are connected. The range of allowed weights and connections is then limited by the routing topology and neuMOS device size. Note that the control of weights through $C_{ij}$ works best if $C_0 \gg \sum_j C_{ij}$ so that $W_{i,j} \approx C_{i,j} / C_0$, however it is possible to design a weighted p-bit design without this assumption ($C_0 \gg \sum C_{ij}$) as discussed in detail in Ref.~\cite{hassan_voltage-driven_2018}.

Similar control can also be achieved through a network of resistors. The weights are given by the same expression, but with capacitances $C_{ij}$ replaced by conductances $G_{ij}$ \cite{hu2016dot}. However, the input conductance $G_0$ of FET's is typically very low, so that an external conductance has to be added to make $G_0 \gg \sum_j G_{ij}$.

\section{Applications of p-circuits}\label{sec:MLI}
As noted earlier, real applications involve  p-bits interconnected by a synapse that can be implemented  off-chip either in software or with a hardware matrix multiplier, but then it is necessary to transfer data back and forth between Eq.~\ref{eq:pbit} and Eq.~\ref{eq:synapse}. Therefore, a low-level compact hardware implementation of a p-bit along with a local synapse as envisioned in Fig.~\ref{fi:fig4} could be a hardware accelerator for many types of applications, some of which will be discussed in this section. In the capacitvely weighted p-bit design of Fig.~\ref{fi:fig4}, the weights and connectivity of the of the p-bit could be dynamically adjusted based on the encoding of a given problem by leveraging a network of programmable switches\cite{lemieux_design_2004} as would be encountered in FPGAs. Such a p-bit with local interconnections would look like a compact nanodevice implementation of  highly scaled digital spiking neurons of neuromorphic chips such as TrueNorth \cite{merolla2014million}. Alternatively, the interconnection function could be performed off-chip using standard CMOS devices such as FPGAs or GPUs while p-bits are implemented in a standalone chip by modifying embedded MRAM technology. Note however, the off-chip implementation of the interconnection matrix would impose a timing constraint for an  asynchronous mode of operation, which requires the weighted summation operation (Eq.~\ref{eq:synapse}) to operate much faster than the p-bit operation (Eq.~\ref{eq:pbit}) for proper convergence\cite{sutton_intrinsic_2017,pervaiz2017hardware}. A full on-chip implementation of a reconfigurable p-bit could function as a low-power, efficient hardware accelerator for applications in Machine Learning and Quantum Computing, but in the near term a heterogenous multi-chip synapse / p-bit combination could also prove to be useful. 

Now that we have discussed some possible approaches to implementing Eqs.~\ref{eq:pbit} and \ref{eq:synapse} in hardware, let us present a few illustrative \emph{p-bit networks} that can implement useful functions and can be built using existing technology. Unless otherwise stated, these results are obtained from full SPICE simulations\cite{camsari2015modular} that solve the stochastic Landau-Lifshitz-Gilbert equation coupled with the PTM-based transistor models in SPICE\cite{cao2002predictive} to model the embedded MTJ based 3-terminal p-bit described in Fig.~\ref{fi:fig3}. 

\subsection{Applications: Machine learning inspired}\label{sec:MLI}
\begin{figure} [t!]
        \includegraphics[width=.95\linewidth]{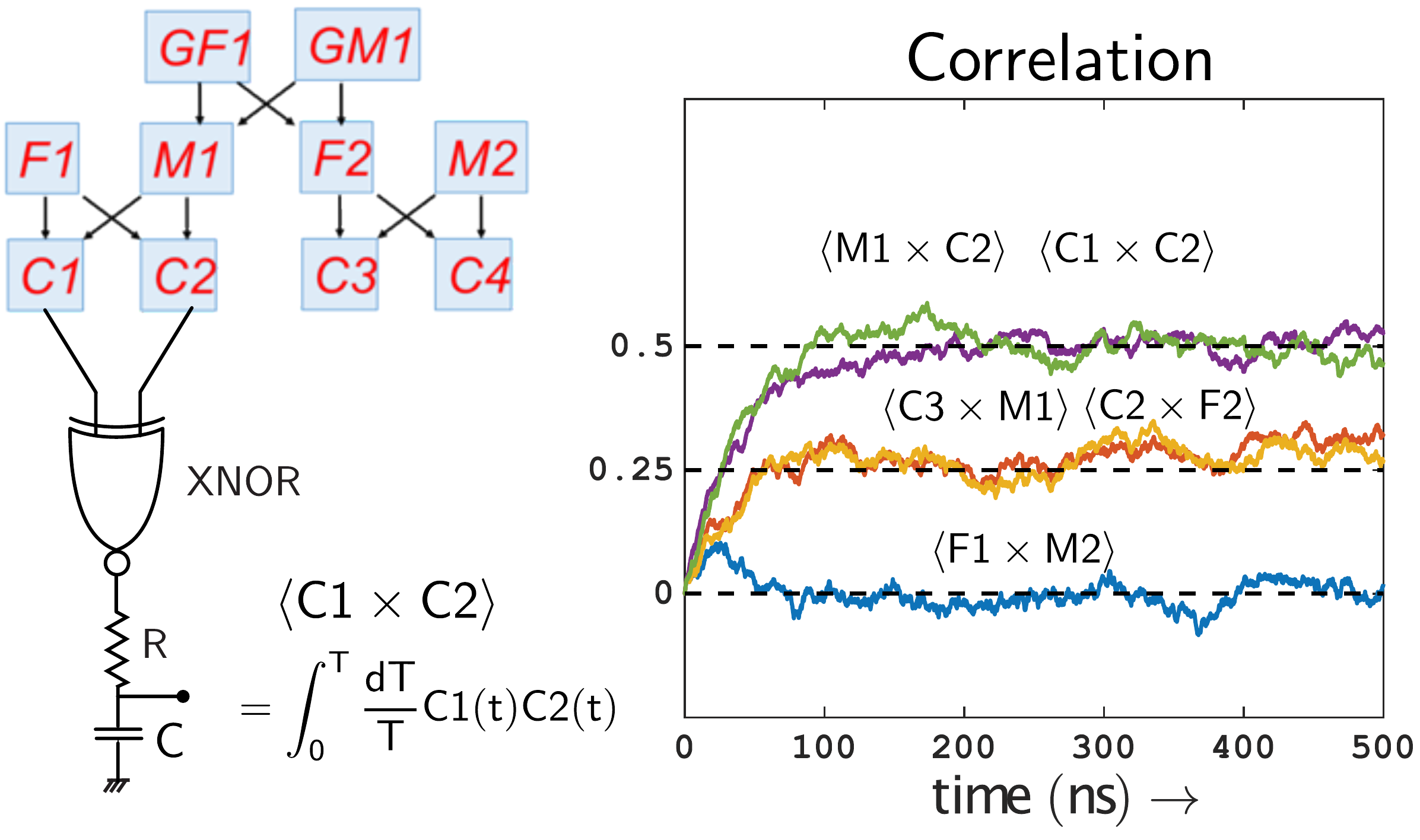}
        \caption{{\bf Genetic circuit:} C1 and C2 are siblings with parents F1, M1, while C3 and C4 are siblings with parents F2, M2. Two of the parents M1 and F2 are siblings with parents GF1, GM1. Genetic correlations between different members can be evaluated from the correlations of the nodal voltages in a p-circuit. An XNOR gate finds their product while a long time constant RC circuit provides the time average.}
                \label{fi:fig5}
\end{figure}

\noindent\textbf{Bayesian inference:} A natural application of stochastic circuits is in the simulation of networks whose nodes are stochastic in nature (See for example \cite{behin-aein_building_2016,chakrapani2007probabilistic,querlioz2015bioinspired,     shim2017stochastic}). An archetypal example is a genetic network, a small version of which is shown in Fig.~\ref{fi:fig5}. A well-known concept is that of genetic correlation or relatedness between different members of a family tree. For example, assuming that each of the children C$_1$ and C$_2$ get half their genes from their parents F$_1$ and M$_1$ we can write their correlation as:
 \begin{eqnarray}
   && \langle C_1\times C_2 \rangle = \langle (0.5 F_1 + 0.5  M_1) \times (0.5 F_1 + 0.5 M_1) \rangle \nonumber \\
   &=& \frac{1}{4} ( \langle F_1 \times F_1 \rangle + \langle F_1 \times M_1 \rangle + \langle M_1 \times F_1 \rangle + \langle M_1 \times M_1 \rangle \nonumber \\
   &=& \frac{1}{4} (1 + 0 + 0 + 1) = 0.5 
\end{eqnarray}
\noindent  \normalsize assuming $\rm F_1$ and $\rm M_1$ are uncorrelated. Hence the well-known result that siblings have 50\% relatedness. Similarly one can work out the relatedness of more distant relationships like that of an aunt $\rm M_1$ and her nephew $\rm C_3$ which turns out to be 25\%.

The point is that we could construct a p-circuit with each of the nodes represented by a hardware p-bit interconnected to reflect the genetic influences. The correlation between two nodes, say C$_1$ and C$_2$, is given by
\begin{align}
\rm \langle C_1 \times C_2 \rangle = \int_{0}^{T} \frac{dt}{T} \ C_1 (t) C_2 (t)
\end{align}
\noindent If C$_1$(t) and C$_2$(t) are binary variables with allowed values of 1 and 0, then they can be multiplied in hardware with an AND gate. If the allowed values are bipolar, $-$1 and +1, then the multiplication can be implemented with an XNOR gate. In either case the average over time can be performed with a long time constant RC-circuit. A few typical results from SPICE simulations are shown in Fig.~\ref{fi:fig5}. The numerical results in Fig.~\ref{fi:fig5} are in good agreement with Bayes theorem even though the circuit operates asynchronously without any sequencers. This is interesting since software simulations of Eqs.~\ref{eq:pbit} and ~\ref{eq:synapse} with directed weights usually require the nodes to be updated from parent to child. Whether this behavior generalizes to larger directed networks is left for future work.

We use this genetic circuit as a simple illustration of the concept of nodal correlations which appear in many other contexts in everyday life. Medical diagnosis\cite{tylman2016real}, for example, involve symptoms such as, say high temperature, which can have multiple origins or parents and one can construct Bayesian networks to determine different causal relationships of interest.\\

\noindent\textbf{Accelerating learning algorithms:} Networks of p-bits could be useful in implementing \emph{inference} networks, where the network weights are trained offline by a learning algorithm in software and the hardware is used to repeatedly perform inference tasks efficiently \cite{ardakani2017vlsi,zand2018low}. 

\begin{figure} [t!]
        \includegraphics[width=0.25\textwidth]{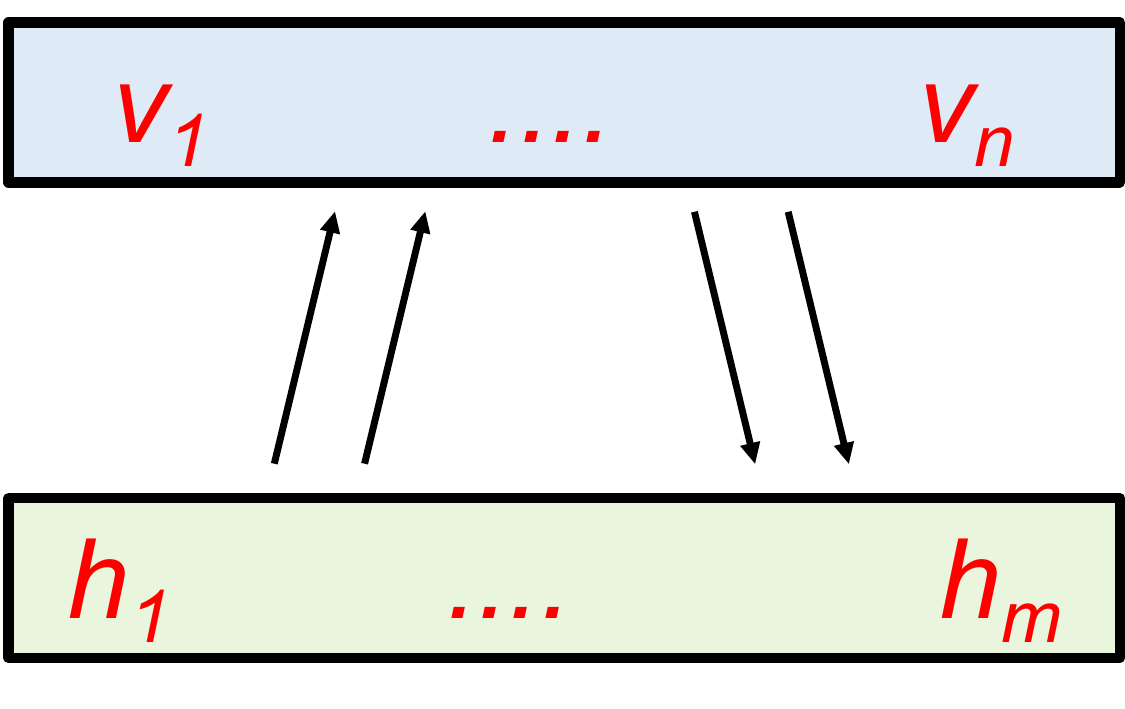}
        \caption{ {\bf Restricted Boltzmann Machine (RBM):} RBMs are a special class of stochastic neural networks that restrict connections within a hidden and a visible layer. Standard learning algorithms require repeated evaluations of correlations of the form $\langle v_i h_j \rangle$.}
        \label{fi:fig6}
    \end{figure}

 Another common example where correlations play an important role is in the learning algorithms used to train modern neural networks like the restricted Boltzmann machine (Fig.~\ref{fi:fig6}) \cite{salakhutdinov2007restricted} having a visible layer and a hidden layer, with connecting weights $W_{ij}$ linking nodes of one layer to those in the other, but not within a layer. A widely used algorithm based on ``contrastive divergence'' \cite{hinton2002training} adjusts each weight $W_{ij}$ according to
\begin{align*}
\Delta W_{ij} \sim \langle v_i h_j \rangle_{t=0} - \langle v_i h_j \rangle_{t\rightarrow \infty} 
\end{align*}
\noindent which requires the repeated evaluation of the correlations $\langle v_i h_j \rangle$. Computing such correlations exactly becomes intractable due to their exponential complexity in the number of neurons\cite{}, therefore contrastive divergence is often limited by a fixed number of steps (CDn) to limit the number of repeated evaluation of these correlations. This process could be accelerated through an efficient physical representation of the neuron and the synapse\cite{bojnordi2016memristive,faria_accelarating_2018}.

\subsection{Applications: Quantum inspired}

The functionality of neural networks is determined by the weight matrix $W_{ij}$ which determines the connectivity among the neurons. They can be classified broadly by the relation between $W_{ij}$ and $W_{ji}$. In traditional feedforward networks, information flow is directed with neuron `i' influencing neuron `j' through a non-zero weight $W_{ij}$ but with no feedback from neuron `j' , such that $W_{ji}=0$. At the other end of the spectrum, is a network with all connections being reciprocal $W_{ij}=W_{ji}.$ In between these two extremes are the class of networks for which the weights between two nodes are asymmetric, but non-zero.

The class of networks with symmetric connections is particularly interesting since they have a close parallel with classical statistical physics where the natural connections between interacting particles is symmetric and the equilibrium probabilities are given by the celebrated Boltzmann law expressing the probability of a particular configuration $\alpha$ in terms of an energy $E_{\alpha}$ associated with that configuration.
\begin{align}
P_{\alpha} = \frac{1}{Z} \ \exp \big( - E_{\alpha} \big) \\
E_{\alpha} = - \{m\}^{T}_{\alpha} \ \ [W]  \ \ \{m\}_{\alpha} \label{eq:energy}
\end{align}

\noindent where $T$ denotes transpose and the constant $Z$ is chosen to ensure that all $P_{\alpha}'s$ add up to one. This energy principle is only available for reciprocal networks \cite{amit1992modeling},  and can be very useful in determining the appropriate weights $W_{ij}$ for a particular problem. 

This class of networks connects naturally to the world of quantum computing which is governed by Hermitian Hamiltonians, and is also the subject of the emerging field of \emph{Ising computing}\cite{yamaoka201620k,mcmahon2016fully,behin-aein_building_2016,sutton_intrinsic_2017,shim2017ising,wang2017oscillator,van2018coherent}. 

\begin{figure} [t]
    \includegraphics[width=0.95\linewidth]{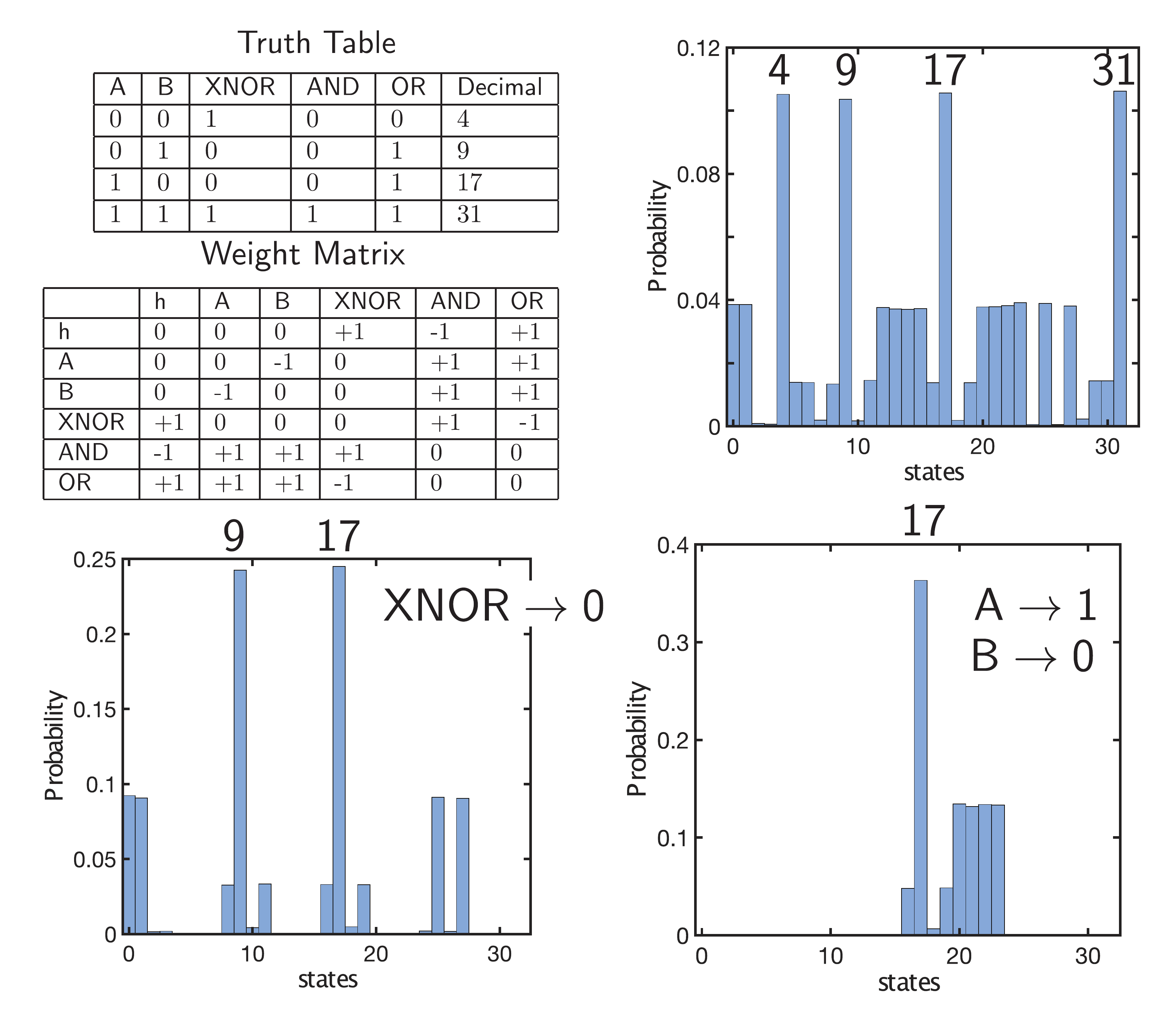}
    \caption{\label{fig:fig9_comb_opt} {\bf Invertible Boolean logic:} A multi-function Boolean gate with 6 p-bits is shown. Inputs A and B produce the output for a 2-input XNOR, AND and OR gate, respectively. The handle bit, ``h'' is used to remove the complementary low-energy states that do not belong to the truth table shown. In the unclamped mode, the system shows the states corresponding to the the lines of the truth table with high probability. A and B can be clamped to produce the correct output for the XNOR, AND and OR in the direct mode. In the inverse mode, any one of the outputs (XNOR is shown as an example) can be clamped to a given value, and the inputs fluctuate among possible input combinations corresponding to this output. }
    \label{fi:fig8}
\end{figure}

\noindent\textbf{Invertible Boolean logic:} Suppose, for example, we wish to design a Boolean gate which will provide three outputs reflecting the AND, OR and XNOR functions of the two inputs A and B. The truth table is shown in Fig.~\ref{fi:fig8}. Note that although we are using the binary notation 1 and 0, they actually stand for p-bit values of +1 and $-$1 respectively.

Since there are five p-bits, two representing the inputs and three representing the outputs, the system has $2^5 = 32$ possible states, which can be indexed by their corresponding decimal values. Each of these configurations has an associated energy, $E_n, n=0,1, \ldots , 31.$ What we need is a weight matrix $W_{ij}$ such that the desired configurations 4, 9, 17 and 31 (in decimal notation) specified by the truth table have a low energy $E_{\alpha}$ (Eq.~(\ref{eq:energy})) compared to the rest, so that they are occupied with higher probability. This can be done either by using the principles of linear algebra \cite{camsari_stochastic_2017} or by using machine learning algorithms \cite{ackley1985learning} to obtain the weight matrix shown in Fig.~\ref{fi:fig8}. Note that an additional p-bit labeled ``h'' has been introduced which is clamped to a value of +1 by applying a large bias.

On the right of Fig.~\ref{fi:fig8}, a histogram is showing the frequency of all the possible (32) configurations obtained from a simulation of Eq.~(\ref{eq:pbit}) and Eq.~(\ref{eq:synapse}) using this weight matrix. Similar results are obtained from a SPICE simulation of a p-circuit of weighted p-bits. Note the peaks at the desired truth table values, with smaller peaks at some of the undesired values. The peaks closely follow the Boltzmann law, such that
\begin{align*}
\frac{P_{\rm desired}}{P_{\rm undesired} }=\exp \big(E_{\rm undesired}-E_{\rm desired} \big)
\end{align*}
\noindent Undesired peaks can be suppressed if we make the W-matrix larger, say by an overall multiplicative factor of 2. If all energies are increased by a factor of 2, the ratio of probabilities would be squared: a ratio of 10 would become a ratio of 100.

It is also possible to operate the gate in a traditional feed-forward manner where inputs are specified and an output is obtained. This mode is shown in the middle panel on the right where the inputs A and B are clamped to 1 and 0 respectively. Only one of the four truth table peaks can be seen, namely the line corresponding to A=1, B=0, which is labeled 17.

What is more interesting is that the gates can be run in \textit{inverse mode} as shown in the lower right panel. The XNOR output is clamped to 0 corresponding to specific lines of the truth table corresponding to 9 and 17. The inputs now fluctuate between the two possibilities, indicating that we can use these gates to provide us with all possible inputs consistent with a specified output, a mode of operation not possible with standard Boolean gates.

This invertible mode is particularly interesting because there are many cases where the direct problem is relatively easy compared to the inverse problem. For example, we can find a suitable weight matrix to implement an adder that provides the sum S of numbers A, B and C. But the same network also solves the inverse problem where a sum S is provided and it finds combinations of k numbers that add up to S\cite{hassan_voltage-driven_2018,pervaiz2018weighted}. This inverse k-sum or subset sum problem is known to be NP-complete\cite{murty1987some} and is clearly much more difficult than direct addition. Similarly we can design a weight matrix such that the network multiplies any two numbers. In inverse mode the same network can factorize a given number, which is a hard problem\cite{shor1999polynomial}. This ability to factorize has been shown with relatively small numbers\cite{camsari_stochastic_2017,pervaiz2018weighted}. How well p-circuits will scale to larger factorization problems remains to be explored. 

 It is worth mentioning that this method of solving integer factorization and the subset sum problem is similar to the deterministic ``memcomputing'' framework where a ``self-organizing logic circuit'' is set up to solve the direct problem and operated in reverse to solve the inverse problem (See for example, Ref. \cite{traversa2017polynomial,diVentra2}). \\
 
 \begin{figure} [t!]
    \includegraphics[width=0.95\linewidth]{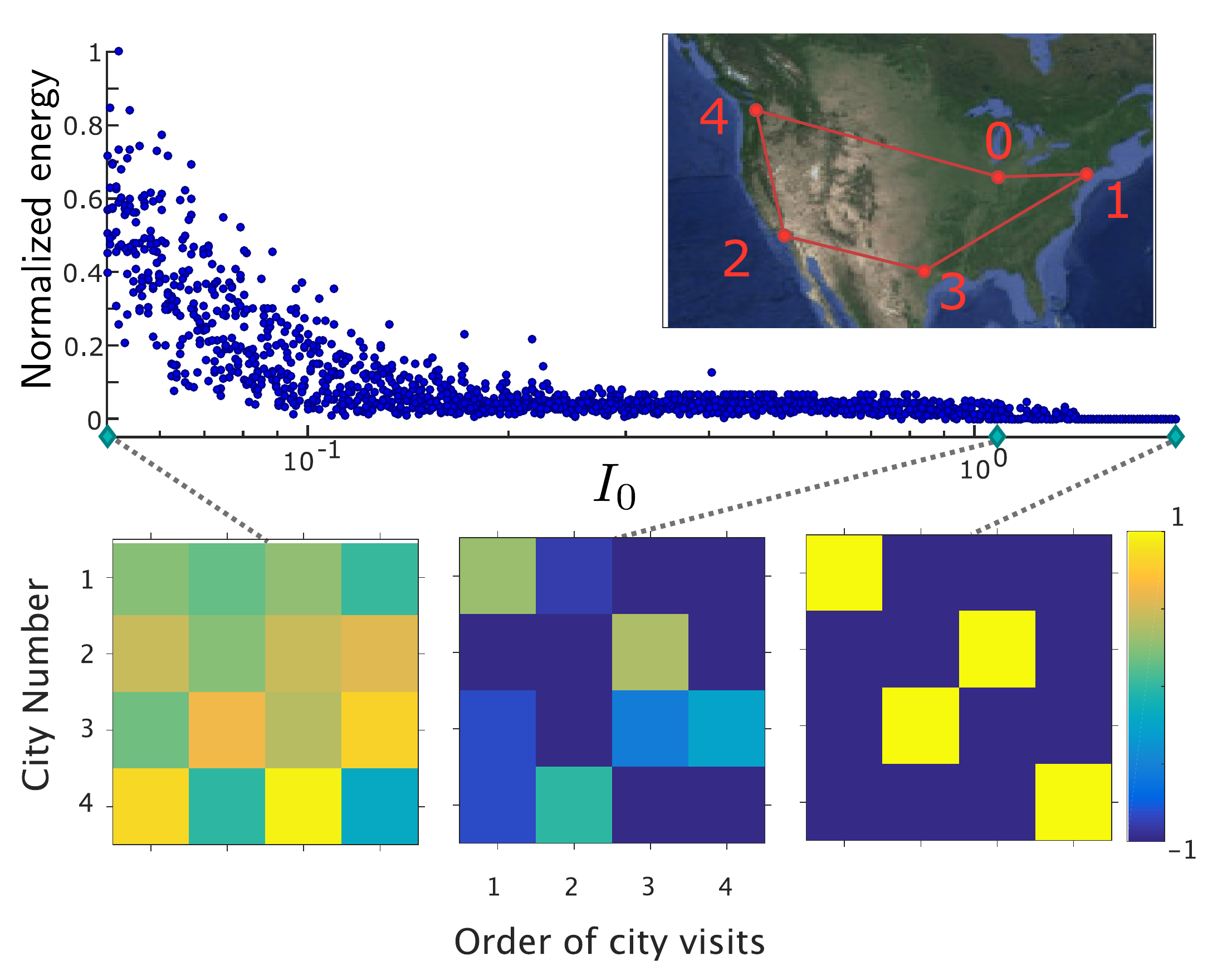}
    \caption{ {\bf Combinatorial Optimization:} A 5-city Traveling Salesman Problem (TSP) implemented using a network of 16 p-bits (fixing city 0), each having two indices, the first denoting the order in which a city is visited and the second denoting the city. The interaction parameter $\sf I_0$ scales all weights and acts as an inverse temperature and is slowly increased via a simple annealing schedule $\sf I_0(t+t_{eq})=(1/0.99) I_0 (t)$ to guide the system into the lowest energy state, providing the shortest traveling distance  (Map imagery data: Google, TerraMetrics).}
    \label{fi:fig9}
\end{figure}

\noindent\textbf{Optimization by classical annealing:} It has been shown that many optimization problems can be mapped onto a network of classical spins with an appropriate weight matrix, such that the optimal solution corresponds to the configuration with the lowest energy \cite{lucas_2014}. Indeed, even the problem of integer factorization discussed above in terms of inverse multiplication can alternatively be addressed in this framework by casting it as an optimization problem\cite{peng2008quantum,henelius_statistical_2011,jiang2018quantum}.

A well-known example of an optimization problem is the classic N-city traveling salesman problem (TSP). It involves finding the shortest route by which a salesman can visit all cities once starting from a particular one. This problem has been mapped to a network of $(N-1)^2$  spins where each spin has two indices, the first denoting the order in which a city is visited and the second denoting the city.

Fig.~\ref{fi:fig9} shows a 5-city TSP mapped to a 16 p-bit network and translated into a p-circuit that is simulated using SPICE. The overall W-matrix is slowly increased and with increasing interaction the network gradually settles from a random state into a low energy state. This process is often called \textit{simulated annealing}\cite{kirkpatrick1983optimization} based on the similarity with the freezing of a liquid into a solid with a lowering of temperature in the physical world, which reduces the random thermal energy relative to a fixed interaction. 

Note that at high values of interaction the p-bits settle to the correct solution with four p-bits highlighted corresponding to (1,1), (2,3), (3,2) and (4,4), showing that the cities should be visited in the order 1-3-2-4. Unfortunately things may not work quite so smoothly as we scale up to problems with larger numbers of p-bits. The system tends to get stuck in metastable states just as in the physical world solids develop defects that keep them from reaching the lowest energy state.\\

\begin{figure} [t!]
    \includegraphics[width=\linewidth]{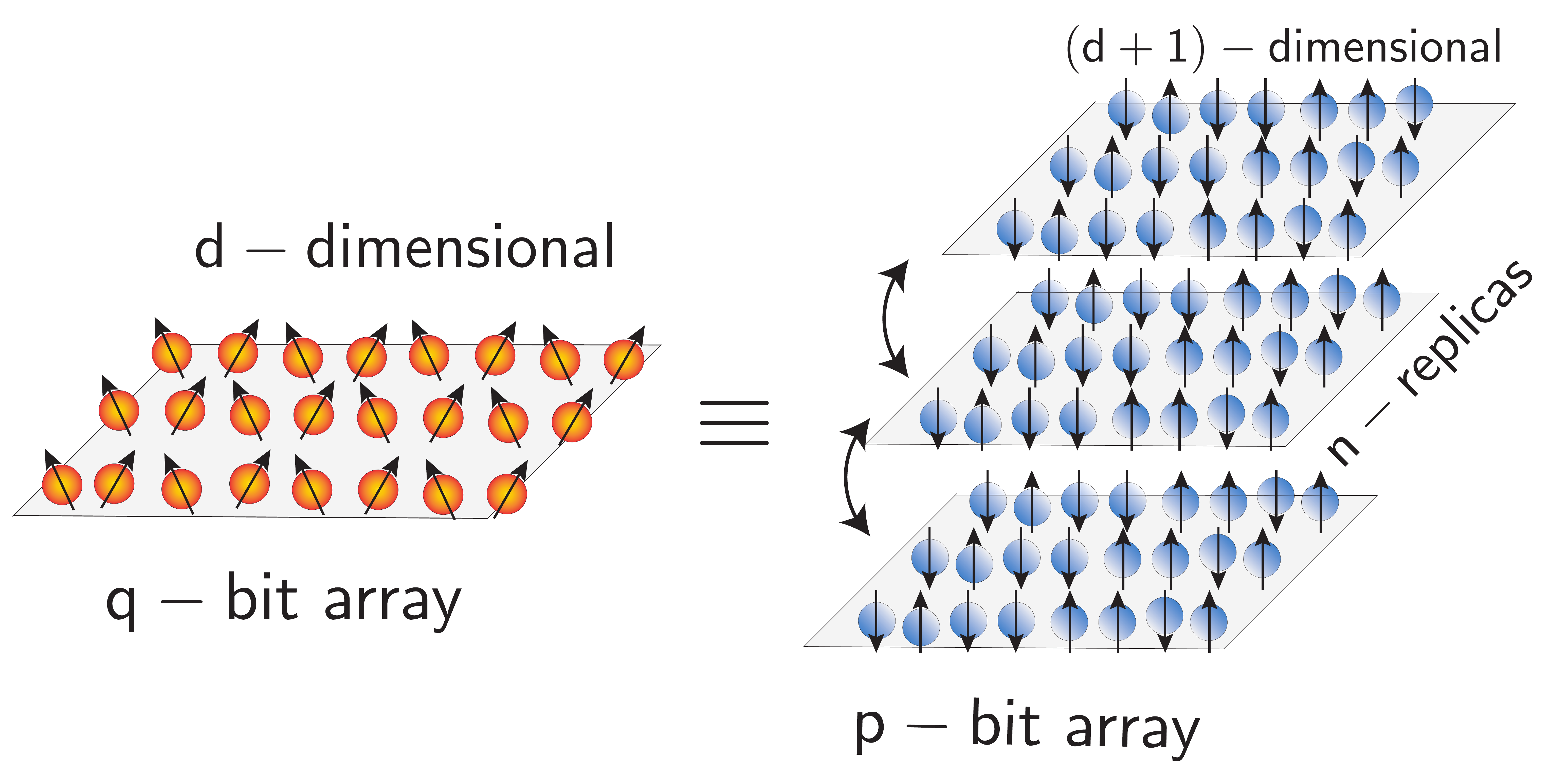}
    \caption{ {\bf Mapping a q-bit network into a p-bit network}: A special class of quantum many body Hamiltonians that are ``stoquastic'' can be solved by mapping them to a classical network of p-bits that consist of a finite number of replicas of the original system that are interacting in the ``vertical'' direction. This approach implemented in software is also known as the Path Integral Monte Carlo method. A hardware implementation would constitute a p-computer that is capable of performing quantum annealing \cite{camsari2018scaled}. }
    \label{fi:fig10}
\end{figure}

\noindent\textbf{Optimization by quantum annealing:} An approach that has been explored is the process of quantum annealing using a network of quantum spins implemented with superconducting q-bits \cite{mooij1999josephson,johnson2011quantum}. However, it is known that for certain classes of quantum problems classified by ``stoquastic'' Hamiltonians \cite{lidar_adiabatic_2018}, a network of q-bits can be approximated with a larger network of p-bits operating in hardware (Fig.~\ref{fi:fig10}) \cite{camsari2018scaled}. We have made use of this equivalence to design p-circuits whose SPICE simulations show correlations and averages comparable to those obtained with quantum annealers \cite{camsari2018scaled}.



\section{Conclusions}

In summary, we have introduced the concept of a probabilistic or p-bit, intermediate between the standard bits of digital electronics and the emerging q-bits of quantum computing. Low barrier magnets or LBM's provide a natural physical representation for p-bits and can be built either from perpendicular magnets (PMA) designed to be close to the in-plane transition or from circular in-plane magnets (IMA). Magnetic tunnel junctions (MTJ) built using LBM's as free layers can be 
combined with standard NMOS transistors to provide three-terminal building blocks for large scale probabilistic circuits that can be designed to perform useful functions. Interestingly, this three-terminal unit looks just like the 1T/MTJ device used in embedded MRAM technology, with only one difference: the use of an LBM for the MTJ free layer.
We hope that this concept will help open up new application spaces for this emerging technology. However, a p-bit need not involve an MTJ, any fluctuating resistor could be combined with a transistor to implement it. It may be interesting to look for resistors that can fluctuate faster based on entities like natural and synthetic antiferromagnets\cite{camsari2016ultrafast,atxitia2018superparamagnetic}, for example.

The p-bit also provides a conceptual bridge between two active but disjoint fields of research, namely stochastic machine learning and quantum computing. This viewpoint suggests two broad classes of applications for p-bit networks. First, there are the applications that are based on the similarity of a p-bit to the binary stochastic neuron (BSN), a well-known concept in machine learning. Three-terminal p-bits could provide an efficient hardware accelerator for the BSN. Second, there are the applications that are based on the p-bit being like a poor man's q-bit. We are encouraged by the initial demonstrations based on full SPICE simulations that several optimization problems including quantum annealing are amenable to p-bit implementations which can be scaled up at room temperature using existing technology.

\begin{acknowledgments}
S.D. is grateful to Dr. Behtash Behin-Aein for many stimulating discussions leading up to Ref~\cite{behin-aein_building_2016}.
\end{acknowledgments}
 \section{References}

    \end{document}